\documentclass[letterpaper,twocolumn]{jpsj3}

\usepackage{txfonts}
\usepackage[dvipdfmx]{graphicx}				
\usepackage{physics}							% TeX will automatically convert eps --> pdf in pdflatex		
\usepackage{amsmath}
\usepackage{amssymb}
\usepackage{braket}
\usepackage{ulem}
\usepackage{bm}
\usepackage{color}
\usepackage{comment}
\usepackage{dcolumn}% Align table columns on decimal point
\usepackage{bbm}% bold math

\newcommand{\bk}{ {\bm{k}} }
\newcommand{\bp}{ \bm{p} }
\newcommand{\bq}{ \bm{q} }
\newcommand{\bH}{ \bm{H} }

\newcommand{\br}{ \bm{r} }

\title{Successive single- and double-$q$ $\Gamma_3$ quadrupole orders on the diamond structure}

\author{Kaito Sasa and Kazumasa Hattori}
\inst{Department of Physics, Tokyo Metropolitan University, 1-1 Minami-Osawa, Hachioji, Tokyo 192-0397, Japan}

\abst{
We investigate field-dependent quadrupolar ordering at the L points of the diamond-structure  Brillouin zone using classical Monte Carlo simulations of an effective $\Gamma_3$ quadrupole model. We demonstrate that competition between the magnetic field and quadrupole anisotropy leads to a rich phase diagram for magnetic fields applied parallel to $[001]$, which includes switching between a single-$\bq$ state and a double-$\bq$ state.
We also examine the effects of a symmetry-allowed biquadratic intersite interaction, corresponding to a hexadecapole interaction, on the stability of the double-$\bq$ state under a magnetic field.
}

\begin{document}
\maketitle

\section{Introduction}

Orbital and multipolar degrees of freedom constitute central themes in strongly correlated electron systems \cite{TokuraNagaosa2000}. In transition-metal oxides, orbital degeneracy and superexchange interactions give rise to orbital ordering and intertwined spin–orbital phenomena \cite{Kugel1973,Khaliullin2005}. From a symmetry-based perspective, orbital order corresponds to the spontaneous ordering of multipole moments transforming according to irreducible representations of the local point group. This viewpoint naturally extends to 5d and f-electron systems \cite{Kim2008-vi,Jackeli2009-lp,Kuramoto2009-ll}, where strong spin–orbit coupling and crystalline electric fields (CEF) generate higher-rank multipoles as active low-energy degrees of freedom \cite{Hayami2018-zx}.

Pr-based cubic 1–2–20 compounds, Pr$T_2X_{20}$ ($T$ = Ti, V, Rh, Ir; $X$ = Al, Zn), provide prototypical realizations of such multipolar physics \cite{OnimaruKusunose2016Review}. In many members of this family, the CEF ground state of the Pr$^{3+}$ ion is a non-Kramers doublet, which carries $\Gamma_3$ electric quadrupole moments $(O_{20},O_{22})\sim \{2z^2-x^2-y^2,\sqrt{3}(x^2-y^2)\}$ under the local point group $T_d$, as well as a $T_{xyz}$ octupole moment, but no magnetic dipole moment \cite{Iwasa2013CEF}. The quadrupole and octupole moments thus constitute the active low-energy multipolar degrees of freedom. Multipolar ordering has been experimentally established in several compounds, including ferroquadrupolar order in PrTi$_2$Al$_{20}$ \cite{Kittaka2020PrTi2Al20,Taniguchi2019PrTi2Al20Switching}, antiferroquadrupolar order in PrIr$_2$Zn$_{20}$ \cite{Onimaru2011PrIr2Zn20,Ishii2011PrIr2Zn20,Iwasa2017PrIr2Zn20Multipoles}, and two-stage quadrupolar–octupolar ordering in PrV$_2$Al$_{20}$ \cite{Sakai2011,Patri2019-fq,Ye2024-vd}. These multipolar ordered states have also been investigated theoretically using mean-field theory \cite{HattoriTsunetsugu2014,Ishitobi2021}, Monte Carlo simulations \cite{HattoriTsunetsugu2016,Freyer2018TwoStage,Freyer2020ThermalFieldTransitions}, impurity models \cite{Kusunose2016-sd}, and Landau theory \cite{Lee2018Landau}.

Beyond static ordering, $\Gamma_3$ systems also exhibit strong quadrupolar fluctuations and unconventional metallic behavior. In PrV$_2$Al$_{20}$, non-Fermi-liquid behavior and strange-metal transport have been reported \cite{Shimura2015-wm,Worl2022Gruneisen,Lenk2024StrangeMetal}. At low temperatures inside the multipolar phase, these compounds also exhibit superconductivity \cite{Onimaru2011PrIr2Zn20,Matsubayashi2012-zg}.  
Theoretical studies have proposed two-channel Kondo lattice scenarios to describe the interplay between quadrupole moments and conduction electrons \cite{TsurutaMiyake2015,InuiMotome2020}, while the microscopic mechanism  of superconductivity remains unclear. These developments underscore the importance of understanding multipolar interactions \cite{Kubo2017-la}.

Among many 1--2--20 compounds, PrIr$_2$Zn$_{20}$ is particularly intriguing.
An antiferroquadrupolar transition occurs at $T_Q \simeq 0.11$~K,
followed by superconductivity at lower temperatures
\cite{Onimaru2011PrIr2Zn20}.
Neutron diffraction measurements have identified the ordering wavevector
at the L point of the Brillouin zone and found a staggered quadrupolar order with $O_{22}$ components on the diamond network
\cite{Iwasa2017PrIr2Zn20Multipoles}.
Because the diamond structure possesses four symmetry-related L points,
competition among multiple ordering wavevectors is intrinsic to the system.

A remarkable feature emerges under an applied magnetic field.
For $\bm{H} \parallel [001]$,
specific-heat measurements reveal two successive anomalies
within a narrow field window
\cite{Onimaru2011PrIr2Zn20,Ishii2011PrIr2Zn20}.
Field-orientation studies further demonstrate pronounced anisotropy
in the phase diagram
\cite{Kittaka2024PrIr2Zn20Pocket}.
Phenomenological Landau analyses have suggested that
competition between single-$\bq$ and double-$\bq$ states
may account for the two-stage transition
\cite{Okanoya2026}.
However, the order parameter, its ordering wavevector, and the microscopic mechanism remain unclear.

In this work, we address this issue by performing
classical Monte Carlo simulations
of an effective $\Gamma_3$ quadrupole model
on the diamond structure 
\cite{HattoriTsunetsugu2016,Freyer2018TwoStage}, focusing on the symmetry-breaking physics while neglecting the $T_{xyz}$ octupole moment and conduction electrons as a zeroth-order approximation. Assuming the ordering wavevector is at the L points as observed for magnetic field along the [110] direction, we demonstrate successive transitions from a paramagnetic phase
to a single-$\bq$ state and subsequently to a double-$\bq$ state,
construct the magnetic-field--temperature phase diagram,
and clarify how higher-order multipolar interactions
modify the competition among L-point quadrupole modes.

This paper is organized as follows.
In Sec.~\ref{sec:Model}, we introduce the effective classical model,
formulated as an XY model on the diamond structure 
with single-ion anisotropy characteristic of $\Gamma_3$ quadrupole systems.
In Sec.~\ref{sec:MC}, we present the Monte Carlo results
and analyze detailed multiple-$\bq$ structure factors.
In the latter part of Sec.~\ref{sec:MC},
we examine the effects of nearest-neighbor biquadratic interactions.
Finally, Sec.~\ref{sec:Discussion} summarizes the mechanism
underlying the phase diagram obtained from the simulations
and compares our results with experimental observations.

%====================================================
\section{Model} \label{sec:Model}
%====================================================

In this paper, we consider an effective model for $\Gamma_3$ quadrupole moments on the diamond structure. 
The two-component $\Gamma_3$ quadrupole moment is represented by a planar vector degree of freedom 
$\bm{m}_i=(m_{xi},m_{yi})\propto (O_{20},O_{22})$ at site $i$ with the fixed-length constraint $|\bm{m}_i|=1$. 
We apply a magnetic field $\bm{H}=(H_x,H_y,H_z)$  within the plane spanned by the [001] and [110] directions. In this geometry, the field–quadrupole coupling takes the form $-h m_{ix}$, where $h=\alpha(2H_z^2-H_x^2-H_y^2)$ with $\alpha>0$.~\cite{HattoriTsunetsugu2014}  In addition, cubic symmetry allows a unique single-ion anisotropy 
$V_i=-b (m_{ix}^3-3 m_{ix} m_{iy}^2)$ \cite{HattoriTsunetsugu2014}. 
Using the parametrization $\bm{m}_i=(\cos\theta_i,\sin\theta_i)$, the Hamiltonian is given by
\begin{equation}
\mathcal{H}
=\sum_{i,j}\Big[J_{ij}\cos\theta_{ij}+K_{ij}\cos^2\theta_{ij}\Big]
-\sum_i \left(h\cos\theta_i + b\cos3\theta_i\right),
\label{eq:H}
\end{equation}
where $\theta_{ij}\equiv\theta_i-\theta_j$, and $J_{ij}$ ($K_{ij}$) denotes the bilinear (biquadratic) exchange interaction on the $ij$ bond. The summation for $ij$ counts each bond once.  
We include exchange interactions up to the fourth neighbor, $J_{1,2,3,4}$. 
To stabilize L-point quadrupolar orders at wavevectors 
$\bm{q}=\bm{k}_1=(\pi,\pi,\pi)$, 
$\bm{k}_2=(-\pi,\pi,\pi)$, 
$\bm{k}_3=(-\pi,-\pi,\pi)$, and 
$\bm{k}_4=(\pi,-\pi,\pi)$, 
a positive $J_4$ is essential, whereas $J_3$ does not contribute to their  stabilization. Our analysis assumes that quadrupolar ordering occurs at the L point wavevectors in  PrIr$_2$Zn$_{20}$. Other ordering wavevectors would require different choices of exchange parameters. 

For magnetic Heisenberg models on the diamond structure, various incommensurate orders are known to appear \cite{Bergman2007,Lee2008-so}, partly due to the geometric frustration inherent in the fcc sublattices of the diamond structure \cite{Henley1987,Gvozdikova_2005,Balla2020-jp}. 
Such incommensurate ordering tendencies can be suppressed by $J_4$, i.e., the second-neighbor interaction within each fcc sublattice \cite{Henley1987}. 
Following the previous study\cite{Okanoya2026}, we set $J_1=1>0$ as the unit of energy and take $J_3=0$ for simplicity. 
For the biquadratic interaction, which corresponds to the hexadecapole--hexadecapole coupling between f-electron moments, we retain only the nearest-neighbor term, denoted simply by $K$. 

The Hamiltonian (\ref{eq:H}) is analyzed by classical Monte Carlo simulations employing the Metropolis algorithm combined with parallel tempering \cite{Hukushima1996-xo}. 
We consider a cubic cluster of linear size $L$ with periodic boundary conditions along the $x$, $y$, and $z$ directions. 
The total number of sites is $N_{\rm tot}=8L^3$. 
The primitive unit cell consists of one site at $\bm{r}\in A$ (fcc sublattice) and another at $\bm{r}+(1/4,1/4,1/4)\in B$, where we set the lattice constant to unity. 
For each temperature $T$, we thermalize the system for $5\times10^4$ Monte Carlo steps (MCSs), followed by measurements over approximately $10^6$ MCSs. 
Thermal expectation values of an observable $O$ are evaluated as Monte Carlo averages, denoted by $\langle O\rangle$. 
We also implement global updates $\theta_i\to\theta_i\pm2\pi/3$ and $\theta_i\to-\theta_i$ for all $i$. 
For finite $h$, only the transformation $\theta_i\to-\theta_i$ remains a symmetry operation.

%%%%%%%%%%%%%%%%%%%%%%%%%%%%%%%%%%%%%%%%%%%%%%%%%%%%%%%%%%

\begin{figure}[t!]
\begin{center}
\includegraphics[width=0.45\textwidth]{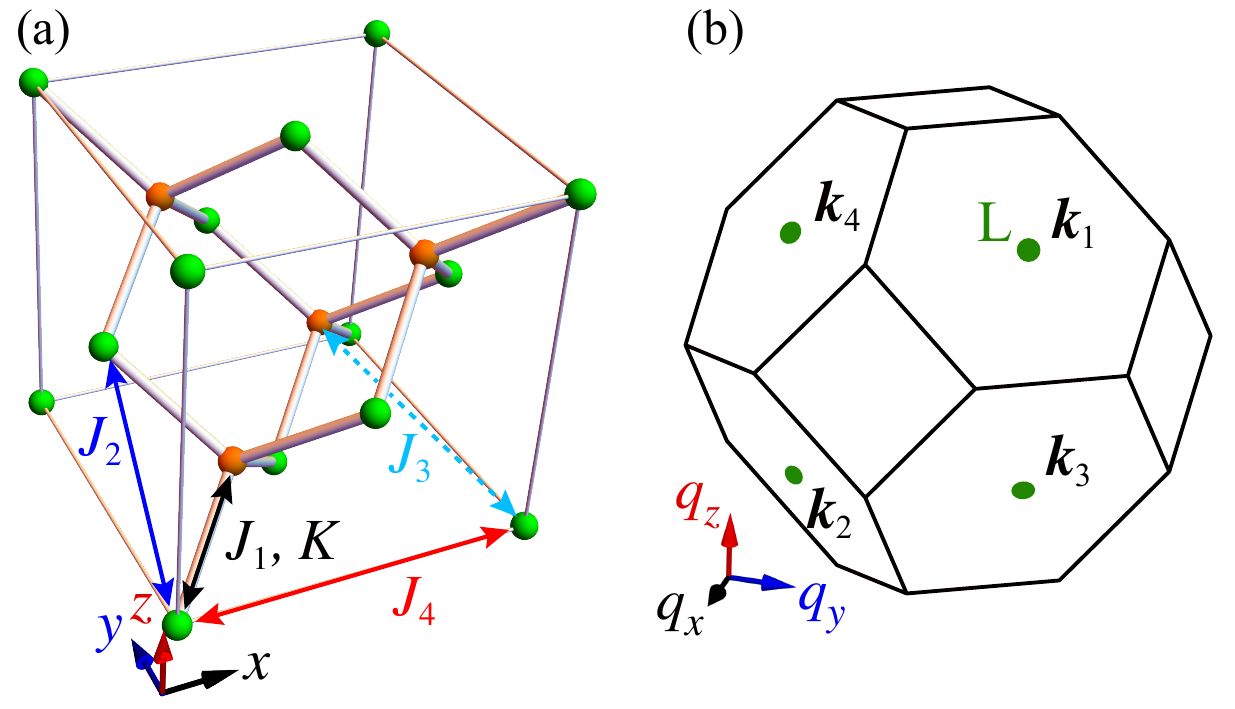}
\end{center}
\caption{ (a) Diamond structure with the sites for the sublattice $A$ ($B$) being indicated in green (orange). Exchange interactions $J_{1,2,3,4}$ and $K$ are indicated. (b) The first Brillouin zone. The L points $\bk_{1,2,3,4}$ are indicated by green circles.}
\label{fig:1}
\end{figure}

%%%%%%%%%%%%%%%%%%%%%%%%%%%%%%%%%%%%%%%%%%%%%%%%%%%%%%%%%% 

\section{Monte Carlo Results} \label{sec:MC}

We now present the results of classical Monte Carlo simulations of the Hamiltonian (\ref{eq:H}). 
We first consider the case with $K=0$ and zero magnetic field $h=0$, 
and subsequently discuss the effects of a finite magnetic field and the biquadratic interaction.

%----------------------------------------------------
\subsection{Zero-field thermodynamics} \label{sec:IIIA}
%----------------------------------------------------

\begin{figure}[t!]
\begin{center}
\includegraphics[width=0.45\textwidth]{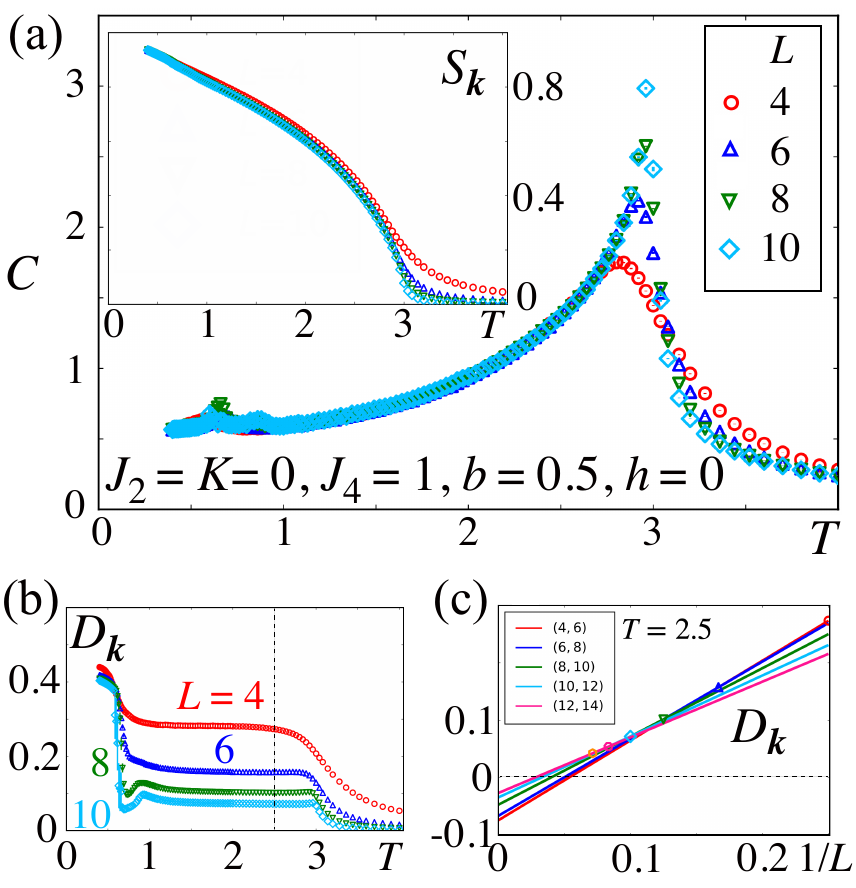}
\end{center}
\caption{(a) Specific heat $C$ as a function of $T$ for $(J_2,J_4,K,h)=(0,1,0,0)$ and $L=4$--$10$. 
 Inset: L-point structure factor $S_{\bk}$. (b) Double-$\bq$ structure factor $D_{\bk}$. (c) Extrapolation of $D_{\bk}$ to $L\to \infty$ for $T=2.5$. See the vertical dashed line in (b). Each straight line is a fit for $L$ and $L+2$ up to $L=12$.}
\label{fig:2}
\end{figure}

Figure~\ref{fig:2}(a) shows the temperature dependence of the specific heat $C$ 
at zero magnetic field ($h=0$) for $J_2=K=0$, $b=0.5$, and $J_4=1$. 
Two distinct anomalies are clearly visible at 
$T_1\simeq 3.010(4)$ and $T_2\simeq 0.7$, where strong finite-size effects prevent a reliable estimate of the uncertainty in $T_2$.  
The higher-temperature peak grows with increasing system size, 
and its position exhibits only weak finite-size dependence, 
indicating a thermodynamic phase transition. 
The lower-temperature anomaly is smaller but becomes sharper 
for larger $L$, suggesting a second phase transition at $T_2$. 

To clarify the nature of these transitions, we compute the L-point  structure factor:
\begin{align}
S_{\bm{k}}
&\equiv \sum_{\ell=1}^4 \Big\langle \big| \bm{m}(\ell) \big|^2 \Big\rangle, \\
\bm{m}(\ell)
&= \frac{1}{N_{\rm tot}} \sum_j p_\ell(j)\,\bm{m}_j 
   e^{i\bm{k}_\ell\cdot\bm{r}_j},
\label{eq:Sk}
\end{align}
where $j$ runs over all sites and $\bm{r}_j$ denotes the position of the corresponding fcc unit-cell origin, defined using the A sublattice. The sublattice-dependent factor is defined as 
\begin{align}
p_\ell(j)=
\begin{cases}
\hspace{2.5mm} 1, & (\ell=1),\\
\hspace{2.5mm} 1, & (\ell\ge 2 \ \text{and}\ j\in A\text{-sublattice}),\\
-1, & (\ell\ge 2 \ \text{and}\ j\in B\text{-sublattice}).
\end{cases}
\label{eq:P}
\end{align}
We also define the double-$\bm{q}$ structure factor as
\begin{align}
D_{\bm{k}}
\equiv \sum_{\ell=1}^4 \sum_{\ell'<\ell}
\Big\langle \big| \bm{m}(\ell) \big| \, \big| \bm{m}(\ell') \big| \Big\rangle.
\end{align}
Note that $D_{\bm{k}}\to 0$ in the thermodynamic limit ($L\to\infty$) 
for any single-$\bm{q}$ ordered state, 
whereas $S_{\bm{k}}$ remains finite for both single-$\bm{q}$ and double-$\bm{q}$ orders. 
Because of the unit-cell convention, $\bm{k}_1$ appears inequivalent in the definition of $p_\ell(j)$ in Eq.~(\ref{eq:P}) compared with the other L points $\bm{k}_{2,3,4}$. 
This apparent asymmetry arises because the direction of $\bm{k}_1$ coincides with the bond direction defining our unit cell. 
We have numerically confirmed that $\langle |\bm{m}(\ell)|^2 \rangle$ is identical for all $\ell$ within numerical accuracy. 
This equivalence can also be understood by explicitly constructing the order-parameter configurations for the four cases, which are related to one another by appropriate symmetry operations. See Appendix \ref{app} for their real-space order parameter configurations.

%%%%%%%%%%%%%%%%%%%%%%%%%%%%%%%%%%%%%%%%%%%%%%%%%%%%%%%%%%

%%%%%%%%%%%%%%%%%%%%%%%%%%%%%%%%%%%%%%%%%%%%%%%%%%%%%%%%%% 

As shown in the inset of Fig.~\ref{fig:2}(a), 
$S_{\bm{k}}$ becomes finite for $T<T_1$, 
showing finite-size behavior consistent with a continuous phase transition. Although cubic invariants are forbidden for the L-point order parameters, the presence of four symmetry-equivalent L-point modes may affect the critical behavior. We leave a detailed analysis of the universality class for future study.

To determine whether the ordered state is of single-$\bm{q}$ type, 
we examine $D_{\bm{k}}$ [Fig.~\ref{fig:2}(b)]. 
For $T<T_2$, the data for $L=4$--$10$ saturate to a finite value. 
We have also inspected Monte Carlo snapshot configurations and find  that 
only two of the L-point components remain macroscopically finite. 
These results indicate that the low-temperature phase ($T<T_2$) 
is a double-$\bm{q}$ ordered state. 

For $T_2<T<T_1$, the situation is more subtle, 
as evidenced by the pronounced finite-size effects in Fig.~\ref{fig:2}(b). 
To analyze these effects, we extend the system size up to $L=14$, 
as shown in Fig.~\ref{fig:2}(c). 
Although the accessible sizes are still limited, 
$D_{\bm{k}}$ decreases monotonically toward zero as $1/L \to 0$. 
We therefore conclude that a single-$\bm{q}$ ordered state 
is realized in the intermediate-temperature regime $T_2<T<T_1$. The large finite-size effect might be related to the degeneracy of the single-$\bq$ and double-$\bq$ orders in the Landau analysis without mode-mode coupling corrections \cite{Okanoya2026}. 
%%%%%%%%%%%%%%%%%%%%%%%%%%%%%%%%%%%%%%%%%%%%%%%%%%%%%%%%%%

\begin{figure}[t!]
\begin{center}
\includegraphics[width=0.45\textwidth]{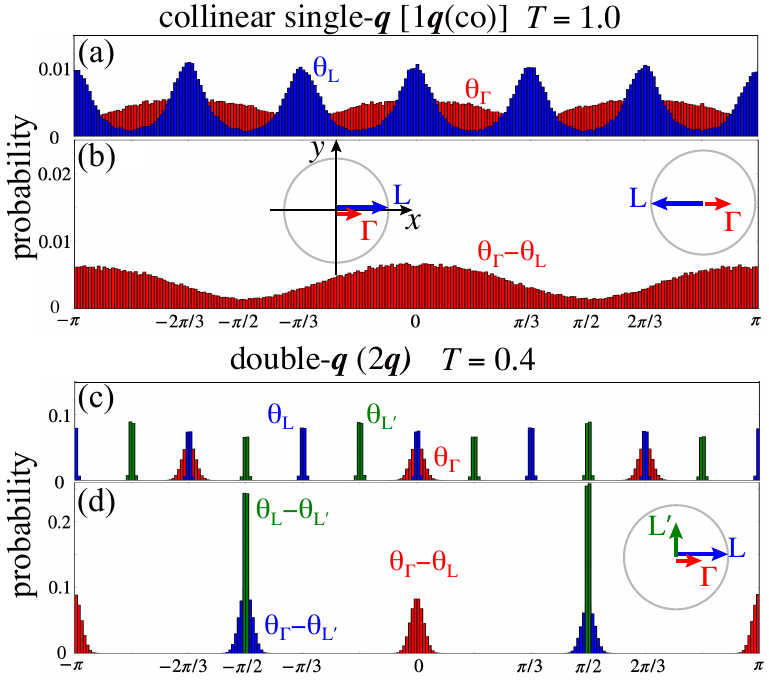}
\end{center}
\caption{Order parameter angle distributions for $b=0.5$, $h=0$, $J_2=K=0$, and $J_4=1.0$. The system size is $L=8$ and $2\times 10^4$ MCSs are used. (a) $\theta_\Gamma$ and $\theta_{\rm L}$ and (b) $\theta_\Gamma-\theta_{\rm L}$ for $T=1.0$, where the collinear single-$\bq$ order is realized. (c) $\theta_\Gamma$, $\theta_{\rm L}$, and $\theta_{\rm L'}$ and (d) $\theta_\Gamma-\theta_{\rm L}$, $\theta_\Gamma-\theta_{\rm L'}$, and  $\theta_{\rm L}-\theta_{\rm L'}$ for $T=0.4$, where the double-$\bq$ order is realized.  In the insets  of panels (b) and (d), schematic configurations of order parameters are illustrated for typical selected domains.}
\label{fig:3}
\end{figure}

%%%%%%%%%%%%%%%%%%%%%%%%%%%%%%%%%%%%%%%%%%%%%%%%%%%%%%%%%% 

Let us discuss the nature of each phase. 
First, in the single-$\bm{q}$ phase, one of the four L-point order parameters 
$\bm{m}(\ell)\equiv\bm{m}_{\rm L}$ becomes finite. 
Through the single-ion anisotropy $V_i$, this ordering induces 
a finite uniform component 
$\bm{m}_\Gamma \equiv N_{\rm tot}^{-1}\sum_j \bm{m}_j$ 
\cite{HattoriTsunetsugu2014}. 
Parameterizing
\begin{align}
\bm{m}_{{\rm L}(\Gamma)}
&=(m_{{\rm L}(\Gamma)x},m_{{\rm L}(\Gamma)y}) \nonumber\\
&=|\bm{m}_{{\rm L}(\Gamma)}|
(\cos\theta_{{\rm L}(\Gamma)},\sin\theta_{{\rm L}(\Gamma)}),
\end{align}
we analyze the angular distributions of the two modes 
$\theta_{\rm L}$ and $\theta_\Gamma$, as shown in Fig.~\ref{fig:3}(a). 
Because the data are taken at relatively high temperatures within the ordered phase, 
the distributions are broad. 
The peak positions of $\theta_{\rm L}$ occur at $\pi n/3$ ($n$: integers), 
while those of $\theta_\Gamma$ appear at $0$ and $\pm 2\pi/3$. 
The distribution of the relative angle 
$\theta_\Gamma-\theta_{\rm L}$ is shown in Fig.~\ref{fig:3}(b), 
demonstrating that the two components are collinear, 
i.e., the peaks occur at $0$ or $\pi$. 
This configuration corresponds to the collinear single-$\bm{q}$ state 
[1$\bm{q}$(co)] obtained for $\bm{H}\parallel[001]$ 
in the Landau analysis of Ref.~\cite{Okanoya2026}. 

Next, in the double-$\bm{q}$ phase (2$\bm{q}$), 
an additional L-point order parameter develops. 
We denote it by 
$\bm{m}(\ell')\equiv\bm{m}_{\rm L'}
=|\bm{m}_{\rm L'}|(\cos\theta_{\rm L'},\sin\theta_{\rm L'})$, 
with $|\bm{m}_{\rm L'}|\le |\bm{m}_{\rm L}|$. 
The corresponding distributions are shown in 
Figs.~\ref{fig:3}(c) and \ref{fig:3}(d). 
The peak positions of $\theta_\Gamma$ remain at 
$0$ and $\pm 2\pi/3$, 
and those of $\theta_{\rm L}$ at $\pi n/3$, 
as in the single-$\bm{q}$ phase. 
For the secondary L-point component, 
the preferred orientations are 
$\theta_{\rm L'}=\pi/2 \pm \pi n/3$. 
Analysis of the relative-angle distribution reveals that 
the two symmetry-related L-point components coexist 
with nearly orthogonal orientations. 
Such orthogonality minimizes the effective quartic coupling 
$[\bm{m}(\ell)\cdot\bm{m}(\ell')]^2$ 
between modes at $\bm{q}_\ell$ and $\bm{q}_{\ell'}$ 
\cite{Okanoya2026}, 
indicating that fluctuation-induced higher-order terms 
play a crucial role in stabilizing the double-$\bm{q}$ state.

%----------------------------------------------------
\subsection{Temperature--magnetic-field phase diagram}
\label{sec:T-H-phase}
%----------------------------------------------------

We next investigate the evolution of the ordered phases under an applied magnetic field. 
We first consider $\bm{H}\parallel[110]$, which corresponds to $h<0$ in our convention. 
Neutron scattering experiments on PrIr$_2$Zn$_{20}$ indicate that 
a single-$\bm{q}$ L-point order with 
$\bm{m}(\ell)=(0,m_{{\rm L}y})$ 
is realized for $\bm{H}\parallel[\bar{1}10]$. 
We examine whether our model reproduces this field-induced ordering pattern. We then turn to the case of $\bm{H}\parallel[001]$, 
corresponding to $h>0$. 
In the remainder of this subsection, 
we elucidate the $h$--$T$ phase diagrams for both field directions 
and compare the resulting phase structures.

%%%%%%%%%%%%%%%%%%%%%%%%%%%%%%%%%%%%%%%%%%%%%%%%%%%%%%%%%

\begin{figure}[t!]
\begin{center}
\includegraphics[width=0.45\textwidth]{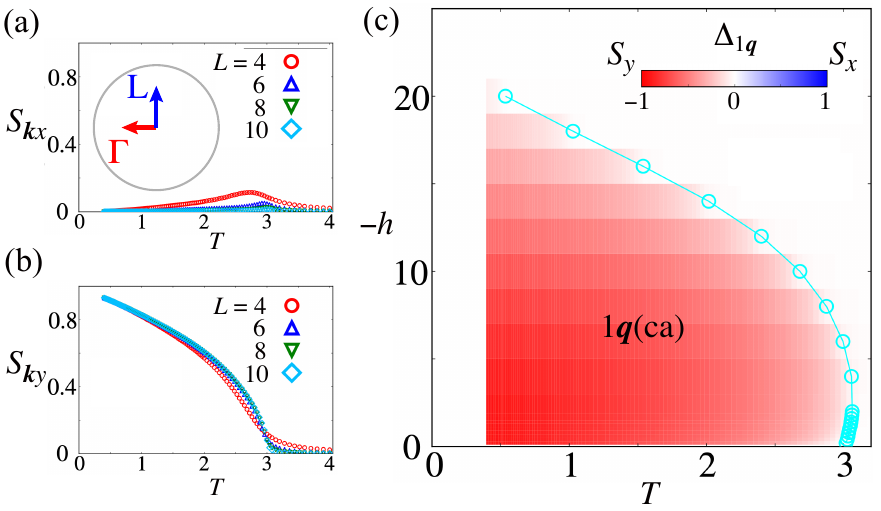}
\end{center}
\caption{ Temperature dependence of the component-resolved structure factors (a) $S_{\bk x}$ and (b) $S_{\bk y}$ for $h=-1.0$, $b=0.5$, $J_2=K=0$, and $J_4=1.0$ with $L=4$, 6, 8, and 10. (c) Phase diagram for $\bH\parallel [110]$ ($h<0$). The color map is $\Delta_{1\bq}=S_{\bk x}-S_{\bk y}$ for $L=8$. The ordering temperature of the $y$ component at the L points is evaluated by the Binder ratio and indicated by circles (cyan). Inset in (a): Schematic order parameter configuration in the canted single-$\bq$ phase labeled as 1$\bq$(ca).}
\label{fig:4}
\end{figure}

%%%%%%%%%%%%%%%%%%%%%%%%%%%%%%%%%%%%%%%%%%%%%%%%%%%%%%%%%% 

\subsubsection{$H \parallel[110]$}
\label{sec:H110}

We consider a magnetic field applied along the [110] direction, 
corresponding to $h<0$ in our convention. 
The pseudo-Zeeman term, $-h\sum_j \cos\theta_j$, 
explicitly breaks the threefold rotational symmetry of the Hamiltonian (\ref{eq:H}). 
This term favors a finite uniform component 
$\bm{m}_\Gamma$ with $\theta_\Gamma=\pi$. 

To distinguish the two components of 
$\bm{m}(\ell)=(m_x(\ell),m_y(\ell))$, 
we introduce anisotropic structure factors,
\begin{align}
S_{\bm{k}x}
&\equiv \sum_{\ell=1}^4 
\Big\langle |m_x(\ell)|^2 \Big\rangle, \ 
S_{\bm{k}y}
\equiv \sum_{\ell=1}^4 
\Big\langle |m_y(\ell)|^2 \Big\rangle.
\label{eq:Skxy}
\end{align}
Figures~\ref{fig:4}(a) and \ref{fig:4}(b) show the temperature dependence of 
$S_{\bm{k}x}$ and $S_{\bm{k}y}$ for $h=-1.0$, 
with the remaining parameters identical to those in Fig.~\ref{fig:2}. 
A clear symmetry breaking is observed in the $S_{\bm{k}y}$ channel, while there is no signature of double-$\bq$ order in $D_\bk$ (not shown). 

The phase diagram for $h<0$ is presented in Fig.~\ref{fig:4}(c), 
where the color scale represents 
$\Delta_{1\bm{q}} \equiv S_{\bm{k}x}-S_{\bm{k}y}$. 
A single-$\bm{q}$ phase with finite $m_y(\ell)$ 
and $\bm{m}_\Gamma$ ($\theta_\Gamma=\pi$) is stabilized. 
In real space, the ordered moments are canted relative to the pseudo-Zeeman field 
along the $x$ direction in $\bm{m}$ space. 
The robustness of this canted single-$\bm{q}$ phase, 
denoted as 1$\bm{q}$(ca) in Fig.~\ref{fig:4}(c), 
originates from the cooperative energy gain of the pseudo-Zeeman 
and single-ion anisotropy terms. 
The leading coupling between the uniform and L-point modes arising from the latter takes the form
$\sim -3b\cos(\theta_\Gamma+2\theta_{\rm L})$, 
which couples the $\Gamma$- and L-point modes \cite{Okanoya2026}. 
Indeed, one finds 
$\theta_\Gamma+2\theta_{\rm L}
=\pi+2(\pm\pi/2)\equiv 0 \ (\mathrm{mod}\ 2\pi)$, 
so that the single-ion anisotropy energy is minimized consistently 
with $\theta_\Gamma=\pi$ favored by the pseudo-Zeeman term. 
For clarity, we do not show results for very small $|h|$, 
where the double-$\bm{q}$ phase persists as in Fig.~\ref{fig:2}. 
Apart from this tiny region of small $|h|$, the single-$\bm{q}$ state is dominant for $\bm{H}\parallel[110]$ for this parameter set.

\subsubsection{$H\parallel[001]$}

\begin{figure}[t!]
\begin{center}
\includegraphics[width=0.47\textwidth]{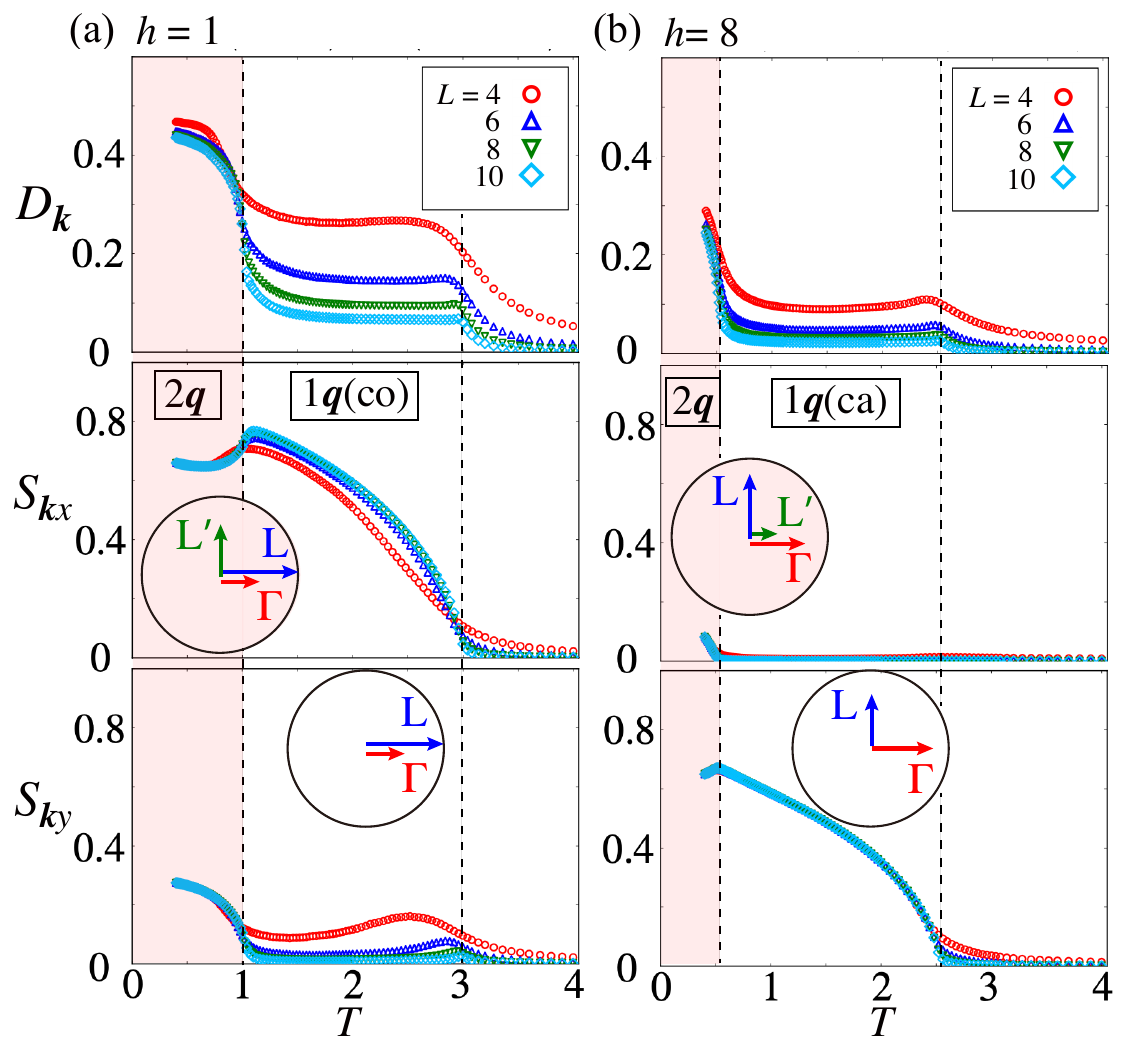}
\end{center}
\caption{The double-$\bq$ structure factor $D_\bk$ (top) and the component-resolved structure factors $S_{\bk x}$ (middle) and $S_{\bk y}$  (bottom) for (a) $h=1$ and (b) $h=8$, with $L=4,6,8$, and 10. The other parameters are the same as those in Fig.~\ref{fig:4}. The vertical dashed lines indicate the phase boundaries as guides for the eye. Schematic order parameter configurations are also shown as in Fig.~\ref{fig:3}. }
\label{fig:5}
\end{figure}

We now discuss the symmetry breaking for $\bm{H}\parallel[001]$ ($h>0$). 
In this case, the pseudo-Zeeman term induces a uniform moment with 
$\theta_\Gamma=0$, opposite in sign to the $\bm{H}\parallel[110]$ case 
discussed in Sec.~\ref{sec:H110}. 
Within a Landau framework, the collinear single-$\bm{q}$ configuration 
incurs a positive quartic term 
$\sim (\bm{m}_\Gamma\cdot\bm{m}_{\rm L})^2$ 
in the free energy. 
Therefore, this state is expected to become unstable at sufficiently large positive $h$, as the uniform component $\bm{m}_\Gamma=(m_{\Gamma x},0)$ increases. Figures~\ref{fig:5}(a) and \ref{fig:5}(b) display 
$D_{\bm{k}}$, $S_{\bm{k}x}$, and $S_{\bm{k}y}$ 
for $h=1$ and $h=8$, respectively, 
with interaction parameters identical to those in Fig.~\ref{fig:4}. 
For the lower field $h=1$ [Fig.~\ref{fig:5}(a)], 
two ordered phases appear: 
the collinear single-$\bm{q}$ state [1$\bm{q}$(co)] 
and the double-$\bm{q}$ (2$\bm{q}$) state, 
similar to the $h=0$ case shown in Figs.~\ref{fig:2} and \ref{fig:3}. 
As discussed in Sec.~\ref{sec:IIIA}, 
the apparent double-$\bm{q}$ signal in the intermediate-temperature region 
$1\lesssim T\lesssim 3$ is attributed to finite-size effects, 
as inferred from the $L$ dependence of $D_{\bm{k}}$ 
in the upper panel of Fig.~\ref{fig:5}(a). 

As $h$ increases, the collinear single-$\bm{q}$ phase disappears, 
and the double-$\bm{q}$ phase is also suppressed, 
as shown in Fig.~\ref{fig:5}(b). 
This suppression arises because configurations with finite 
$m_{{\rm L}x}$ become energetically unfavorable 
as the uniform component $m_{\Gamma x}$ grows. 
Consequently, a canted single-$\bm{q}$ phase [1$\bm{q}$(ca)] 
is stabilized, similar to the $h<0$ case, 
although the single-ion anisotropy term is not fully minimized 
when $m_{\Gamma x}>0$, in contrast to the situation for $h<0$.

%%%%%%%%%%%%%%%%%%%%%%%%%%%%%%%%%%%%%%%%%%%%%%%%%%%%%%%%%%

%%%%%%%%%%%%%%%%%%%%%%%%%%%%%%%%%%%%%%%%%%%%%%%%%%%%%%%%%% 

%%%%%%%%%%%%%%%%%%%%%%%%%%%%%%%%%%%%%%%%%%%%%%%%%%%%%%%%%%

\begin{figure*}[t!]
\begin{center}
\includegraphics[width=\textwidth]{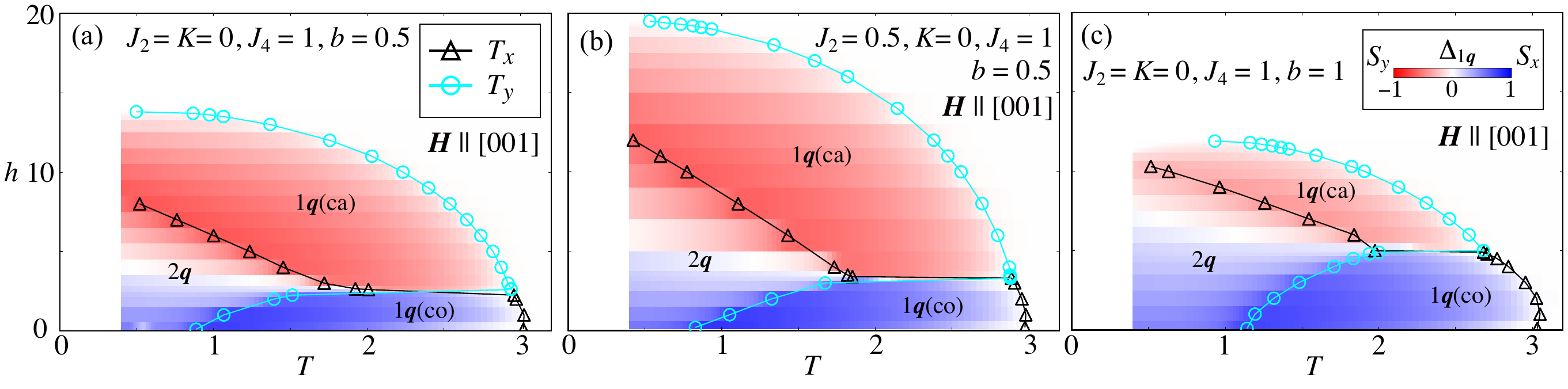}
\vspace{-8mm}
\end{center}
\caption{$T$--$h$ phase diagrams for $J_4=1$, $K=0$, and $\bm{H}\parallel [001]$ ($h>0$). (a) $(J_2,b)$=(0,0.5), (b) $(J_2,b)$=(0.5,0.5), and (c) $(J_2,b)$=(0,1). The color represents $\Delta_{1\bq}=S_{\bk x}-S_{\bk y}$ for $L=8$. The symbols indicate the transition temperatures $T_{\mu}(\mu=x,y)$ at which the $\mu$ component of the L-point order parameter becomes ordered.}
\label{fig:6}
\end{figure*}

%%%%%%%%%%%%%%%%%%%%%%%%%%%%%%%%%%%%%%%%%%%%%%%%%%%%%%%%%% 

We summarize the $T$--$h$ phase diagram for $\bm{H}\parallel[001]$ 
in Fig.~\ref{fig:6}. 
Because the results for $h\le 0$ are qualitatively insensitive 
to microscopic details, 
we have mainly presented data for the simple parameter set with $J_2=0$. 
Here, we examine the effects of $J_2$ and $b$ 
on the phase diagram for $h>0$. 

Figure~\ref{fig:6}(a) shows the phase diagram 
for the parameter set used in Figs.~\ref{fig:2}--\ref{fig:5}, 
whereas Fig.~\ref{fig:6}(b) and Fig.~\ref{fig:6}(c) correspond to 
$(J_2,b)=(0.5,0.5)$ and $(J_2,b)=(0,1)$, respectively. 
In all three cases, the phase diagrams are qualitatively similar, 
consisting of three ordered phases: 
the collinear single-$\bm{q}$ phase in the low-field and high-temperature region, 
the canted single-$\bm{q}$ phase in the high-field region, 
and the double-$\bm{q}$ phase in the low-temperature and low-field region. Quantitatively, a positive $J_2$ suppresses the uniform component, 
thereby increasing the critical field, 
as seen in Fig.~\ref{fig:6}(b) compared with Fig.~\ref{fig:6}(a). 
As illustrated by the colormap in Fig.~\ref{fig:6}, 
the single-ion anisotropy $b$ tends to align the moments 
along the $x$ direction in $\bm{m}$ space for $h>0$. 
Consequently, both the collinear single-$\bm{q}$ 
and the double-$\bm{q}$ phases extend to higher fields, 
as shown in Fig.~\ref{fig:6}(c).

%----------------------------------------------------
\subsection{Effect of the biquadratic interaction}
%----------------------------------------------------

We now investigate the effect of the nearest-neighbor biquadratic interaction 
$K\cos^2\theta_{ij}$. 
To clarify how this term modifies the stability of the symmetry-broken phases discussed above, 
it is convenient to return to the vector representation 
$\bm{m}_i=(m_{xi},m_{yi})$ and rewrite the interaction in Fourier space:
\begin{align}
\sum_{\langle i,j\rangle}\cos^2\theta_{ij}
&\;\to\; \sum_{\langle \br_i,\br_j\rangle}
(\bm{m}_{\br_i}^{A}\cdot\bm{m}_{\br_j}^{B})^2 \nonumber\\
&=\frac{1}{(N_{\rm tot}/2)^3}\sum_{\bm{p},\bm{l},\bm{q}}
\gamma_{\bm{q}}\,
\bm{m}^{A}_{\bm{G}-\bm{q}-\bm{p}}
\cdot
\bm{m}^{B}_{\bm{q}-\bm{l}}\,
\bm{m}^{A}_{\bm{p}}
\cdot
\bm{m}^{B}_{\bm{l}},
\label{eq:H4}
\end{align}
where $\bm{G}$ is a reciprocal lattice vector. We take $i$ and $j$ on the $A$ and $B$ sublattices, respectively, and denote 
$\bm{m}_i\to\bm{m}_{\br_i}^{A}$ and $\bm{m}_j\to\bm{m}_{\br_j}^{B}$ with $\br_{i}$ and $\br_{j}$ being regarded as the corresponding unit cell label, i.e., that for the $A$ sublattice. $\boldsymbol{m}^{A,B}_{\bm p}$ is the Fourier component defined by $\bm{m}^{A,B}_{\br_j}= (N_{\rm tot}/2)^{-1} \sum_{\bp} \,\bm{m}_{\bp}^{A,B} 
   e^{i\bm{p}\cdot\bm{r}_j}
$.
The $\gamma_{\bm{q}}=\sum_{\boldsymbol{\delta}} 
e^{i\bm{q}\cdot\boldsymbol{\delta}}$ 
is the nearest-neighbor form factor with 
$\boldsymbol{\delta}=(0,0,0)$, 
$(-1/2,-1/2,0)$, 
$(0,-1/2,-1/2)$, and 
$(-1/2,0,-1/2)$, where $\boldsymbol{\delta}$ denotes the displacement between the corresponding fcc unit-cell origins.  We retain only  contributions involving the L-point order parameters 
defined in Eq.~(\ref{eq:Sk}). 
There are two classes in which all four $\bm{m}$ fields belong to L-point modes: 
(i) $\bm{q}=\bm{0}$ and 
(ii) $\bm{q}$ at the X points, 
$\bm{q}_{\rm X}=(2\pi,0,0)$, 
$\bm{q}_{\rm Y}=(0,2\pi,0)$, and 
$\bm{q}_{\rm Z}=(0,0,2\pi)$ in Eq.~(\ref{eq:H4}). 
The latter contributions vanish because 
$\gamma_{\bm{q}_{\rm X}}
=\gamma_{\bm{q}_{\rm Y}}
=\gamma_{\bm{q}_{\rm Z}}=0$. 
Therefore, it suffices to consider the $\bm{q}=\bm{0}$ contribution, 
which generates terms of the form 
\begin{align}
\gamma_{\bm{0}}\,
\big(
\bm{m}^{A}_{\bm{k}_\ell}
\cdot
\bm{m}^{B}_{\bm{k}_{\ell'}}
\big)^2,
\qquad (\ell,\ell'=1,2,3,4),
\label{eq:H4correction}
\end{align}
where $\gamma_{\bm{0}}=4$ and we have used $\bm{k}_\ell\equiv -\bm{k}_\ell$ modulo a reciprocal lattice vector. 

Equation~(\ref{eq:H4correction}) generates corrections 
to the quartic terms of Landau free energy, 
\begin{align}
\delta f
\propto
4K \sum_{\ell=1}^4\Bigg\{
|\bm{m}(\ell)|^4
+2
\sum_{\ell'<\ell}
\big[
\bm{m}(\ell)\cdot\bm{m}(\ell')
\big]^2\Bigg\}.
\end{align}
For $K>0$, this contribution favors a double-$\bm{q}$ state 
with mutually orthogonal components 
$\bm{m}(\ell)\perp\bm{m}(\ell')$ 
over a single-$\bm{q}$ state near the transition temperature. 
This can be verified explicitly by substituting, for example, 
$\{\bm{m}(1),\bm{m}(2),\bm{m}(3),\bm{m}(4)\}
=\{\bm{m}_{\rm L},0,0,0\}$ 
for the single-$\bm{q}$ state, and 
$\tfrac{1}{\sqrt{2}}
\{\bm{m}_{\rm L},\bm{m}_{\rm L'},0,0\}$ 
with $\bm{m}_{\rm L}\perp\bm{m}_{\rm L'}$ and 
$|\bm{m}_{\rm L}|=|\bm{m}_{\rm L'}|$ 
for the double-$\bm{q}$ state. In Landau language, a positive $K$ renormalizes the quartic coefficients so as to penalize collinear configurations and favor orthogonal mode coupling, driving the system from single-$\bm{q}$ to double-$\bm{q}$ order.

%%%%%%%%%%%%%%%%%%%%%%%%%%%%%%%%%%%%%%%%%%%%%%%%%%%%%%%%%%

\begin{figure*}[t!]
\begin{center}
\includegraphics[width=\textwidth]{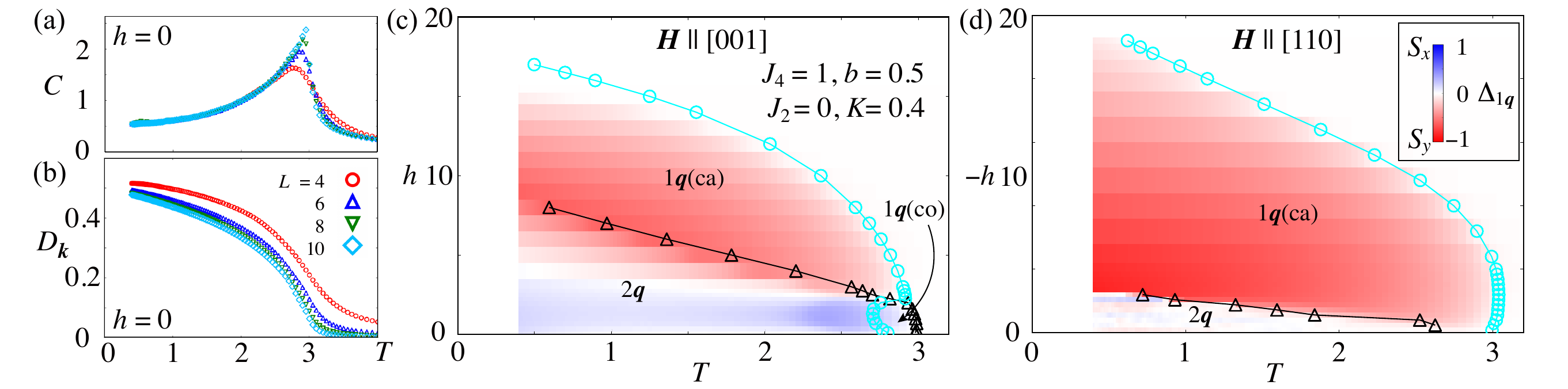}
\vspace{-8mm}
\end{center}
\caption{Temperature dependence of (a) the specific heat $C$ and (b) the double-$\bq$ structure factor $D_\bk$ for $K=0.4$ and $h=0$. $T$--$h$ phase diagrams for $J_2=0$, $J_4=1$, $K=0.4$, and $b=0.5$: (c) $\bm{H}\parallel [001]$ ($h>0$) and (d) $\bm{H}\parallel [110]$ ($h<0$). The color represents $\Delta_{1\bq}=S_{\bk x}-S_{\bk y}$ for $L=8$. The symbols in (c) and (d) indicate the transition temperatures similarly to Fig.~\ref{fig:6}. For $h=0$, the transition temperature $T_2$ for the double-$\bq$ order is slightly lower than $T_1$ for the single-$\bq$, which becomes visible for larger system sizes ($L\ge 14$). }
\label{fig:7}
\end{figure*}

%%%%%%%%%%%%%%%%%%%%%%%%%%%%%%%%%%%%%%%%%%%%%%%%%%%%%%%%%% 

In accordance with the above analysis, the Monte Carlo results reveal that the biquadratic term $K>0$ stabilizes the double-$\bq$ phase. As $K$ increases, the lower transition temperature $T_2$ also increases. At $K\simeq 0.4$, $T_2$ almost merges with $T_1$ at $h=0$. Figures~\ref{fig:7}(a) and \ref{fig:7}(b) show the specific heat $C$ and the double-$\bq$ structure factor $D_\bk$ as functions of $T$ for $h=0$, respectively. There is only a single anomaly in the $T$ dependence of $C$ for $L=10$. We have checked the system size dependence up to $L=16$ and found evidence of a second anomaly associated with $T_2$ separating the single-$\bq$ and double-$\bq$ phases for $K=0.4$.  

Figure~\ref{fig:7}(c) shows the $T$–$h$ phase diagram for $K=0.4$ and $\bH \parallel [001]$. The double-$\bq$ phase indeed expands in the low-field region of the phase diagram. As a result, the collinear phase exists only in the low-$h$ and high-$T$ region. Nevertheless, the collinear phase is stabilized by a small $h$ near the zero-field critical temperature $T_1$. The strong stabilization of the double-$\bq$ phase due to positive $K$ also leads to an expansion of the double-$\bq$ phase for $\bH \parallel [110]$, as shown in Fig.~\ref{fig:7}(d). It should be noted that $\Delta_{1\bq}\simeq 0$ (white) in the double-$\bq$ phase in both Figs.~\ref{fig:7}(c) and \ref{fig:7}(d). This originates from a cancellation between two modes at the L points with $\bm{m}_{\rm L}\parallel (1,0)$ and $\bm{m}_{\rm L'}\parallel (0,1)$. This is in stark contrast to the case of $K=0$ shown in Figs.~\ref{fig:6}(a)–\ref{fig:6}(c).

%====================================================
\section{Discussion} \label{sec:Discussion}
%====================================================

In this study, we have analyzed an anisotropic XY model 
with a cubic single-ion anisotropy characteristic of 
cubic $\Gamma_3$ quadrupole moments. 
Our focus has been on ordering instabilities at the L points, 
which are believed to be realized in PrIr$_2$Zn$_{20}$ 
\cite{Iwasa2017PrIr2Zn20Multipoles}. 
Rather than attempting to map out the full phase diagram 
for general exchange parameters in Eq.~(\ref{eq:H}), 
we have concentrated on the symmetry-breaking properties 
when the leading instability occurs at the L points, 
as motivated by experiments on PrIr$_2$Zn$_{20}$. 

Our results clarify that the successive phase transitions arise 
from competition among symmetry-related L-point components. 
The double-$\bm{q}$ configuration with 
$\bm{m}_{\rm L}\perp\bm{m}_{\rm L'}$ 
reduces the quartic free-energy cost and is therefore stabilized 
at low temperatures. 
This mechanism differs from that proposed for triple-$\bm{q}$ 
quadrupole orders on the fcc lattice 
\cite{Tsunetsugu2021FCC,Hattori2023PartialOrders} 
and from the mechanism of charge-density-wave ordering 
in kagome metals \cite{Denner2021}, where the cubic mode-mode couplings directly couple the modes at the symmetry-related ordering wavevectors. Since our model (\ref{eq:H}) consists of  planar order parameters, the mechanism stabilizing double-$\bq$ order is similar to that discussed for magnetic multiple-q states in models with anisotropic bilinear-biquadratic interactions~\cite{Hayami2020-ac}.

The biquadratic interaction further enhances the stability 
of the double-$\bm{q}$ phase, and we have clarified the underlying mechanism. 
In magnetic Kondo lattice systems, biquadratic interactions 
are known to stabilize multiple-$\bm{q}$ states \cite{Akagi2012-cc,Hayami2014-vz}, 
where higher-order Fermi-surface nesting plays a crucial role. 
For Pr$^{3+}$ ions with a 4f$^2$ configuration whose total angular momentum is $J=4$, 
the biquadratic interaction in terms of the $\Gamma_3$ pseudospin 
corresponds to a hexadecapole--hexadecapole coupling 
and can naturally arise from RKKY processes 
without invoking higher-order nesting. 
Indeed, such terms were introduced in the analysis of 
PrTi$_2$Al$_{20}$ to account for its unusual ferroquadrupolar order 
under magnetic fields \cite{Taniguchi2019PrTi2Al20Switching}. 
In the present study, the biquadratic coupling $K$ 
has been treated phenomenologically; 
it would be an interesting future direction 
to derive such higher-order interactions microscopically 
within the $J=4$ multiplet framework of f-electron systems \cite{Kubo2017-la}.

For comparison with the experimental phase diagram of PrIr$_2$Zn$_{20}$, 
our results predict that the low-field and low-temperature region 
is occupied by a double-$\bm{q}$ phase. 
Experimental verification of this prediction would be highly desirable. 
Upon applying a magnetic field, the double-$\bm{q}$ state is destabilized 
and single-$\bm{q}$ orders emerge. 
For $\bm{H}\parallel[001]$, 
an $O_{20}$ ($m_x$-type) quadrupole order [1$\bq$(co)] is stabilized. 
This phase may become a candidate for the recently identified but previously 
unresolved ``A phase''~\cite{Kittaka2024PrIr2Zn20Pocket}. 
For smaller $K$, clear two transitions appear at zero field, 
inconsistent with experiments 
\cite{Onimaru2011PrIr2Zn20,Kittaka2024PrIr2Zn20Pocket}. 
Because this discrepancy persists across several parameter sets 
with $K=0$, our results suggest that hexadecapolar (biquadratic) interactions deserve further theoretical investigation to clarify the phase diagram of PrIr$_2$Zn$_{20}$, provided that the A phase is characterized by L-point ordering. 
In this sense, previous Landau analyses may have overestimated 
the stability of the double-$\bm{q}$ phase due to contributions generated by integrating out X-point modes. 

For $\bm{H}\parallel[110]$, 
an $O_{22}$ ($m_y$-type) quadrupole order is stabilized, 
consistent with the neutron scattering results 
\cite{Iwasa2017PrIr2Zn20Multipoles}. 
In specific-heat 
\cite{Onimaru2011PrIr2Zn20,Kittaka2024PrIr2Zn20Pocket} 
and thermal-expansion measurements \cite{Okamoto2025PRB}, 
clear signatures of the phase boundaries associated with the low-field 
double-$\bm{q}$ and single-$\bm{q}$ phases have not yet been resolved, 
whereas early ultrasonic experiments reported nearly 
$|\bm{H}|$-independent anomalies \cite{Ishii2011PrIr2Zn20}. 
As for the superconductivity inside the quadrupolar ordered phase \cite{Onimaru2011PrIr2Zn20}, it is interesting to note that this superconductivity occurs inside the double-$\bq$ phase within our proposed phase diagram. This raises the question of whether double-$\bq$ quadrupolar fluctuations can mediate superconducting pairing~\cite{Kubo2018-mi}.

Since PrIr$_2$Zn$_{20}$ is currently the only compound 
for which the multipolar ordering wavevector has been identified, 
it provides a unique platform for studying correlated multipolar 
orders in non-Kramers systems. 
We hope that future experiments will clarify 
the global phase diagram of PrIr$_2$Zn$_{20}$ in addition to the order parameter and the ordering wavevector in the A phase. 

%====================================================
\section{Summary}
%====================================================

We have demonstrated successive single-$\bm{q}$ 
and double-$\bm{q}$ ordering in a quadrupole model using classical Monte Carlo simulations. 
We have elucidated the $T$--$\bm{H}$ phase diagram, 
which partially accounts for existing experimental observations in PrIr$_2$Zn$_{20}$ \cite{Onimaru2011PrIr2Zn20,Ishii2011PrIr2Zn20,Kittaka2024PrIr2Zn20Pocket}
and provides predictions for yet-unconfirmed phase boundaries by extending the previous Landau analysis \cite{Okanoya2026}.  
Our results highlight the importance of the biquadratic interaction, 
corresponding to hexadecapolar coupling, 
in qualitatively reproducing the reported phase diagram. 
We hope that this work stimulates further experimental 
and theoretical studies of multipolar physics 
in PrIr$_2$Zn$_{20}$ and related compounds.

\section*{Acknowledgment}
The authors thank T. Onimaru  and S. Kittaka for fruitful discussions. This work was supported by JSPS KAKENHI 
(Grant No.~JP23K20824, No.~JP23H04866, and No.~JP23H04869).

\appendix

\section{Real-space order parameter configurations} \label{app}
We summarize the real-space order parameter configurations for the phases discussed in the main text. Examples of distinct domain configurations as determined by the ordering wavevectors at the L points are also demonstrated.
\subsection{Single-$q$ orders}
\begin{figure}[t!]
\begin{center}
\includegraphics[width=0.45\textwidth]{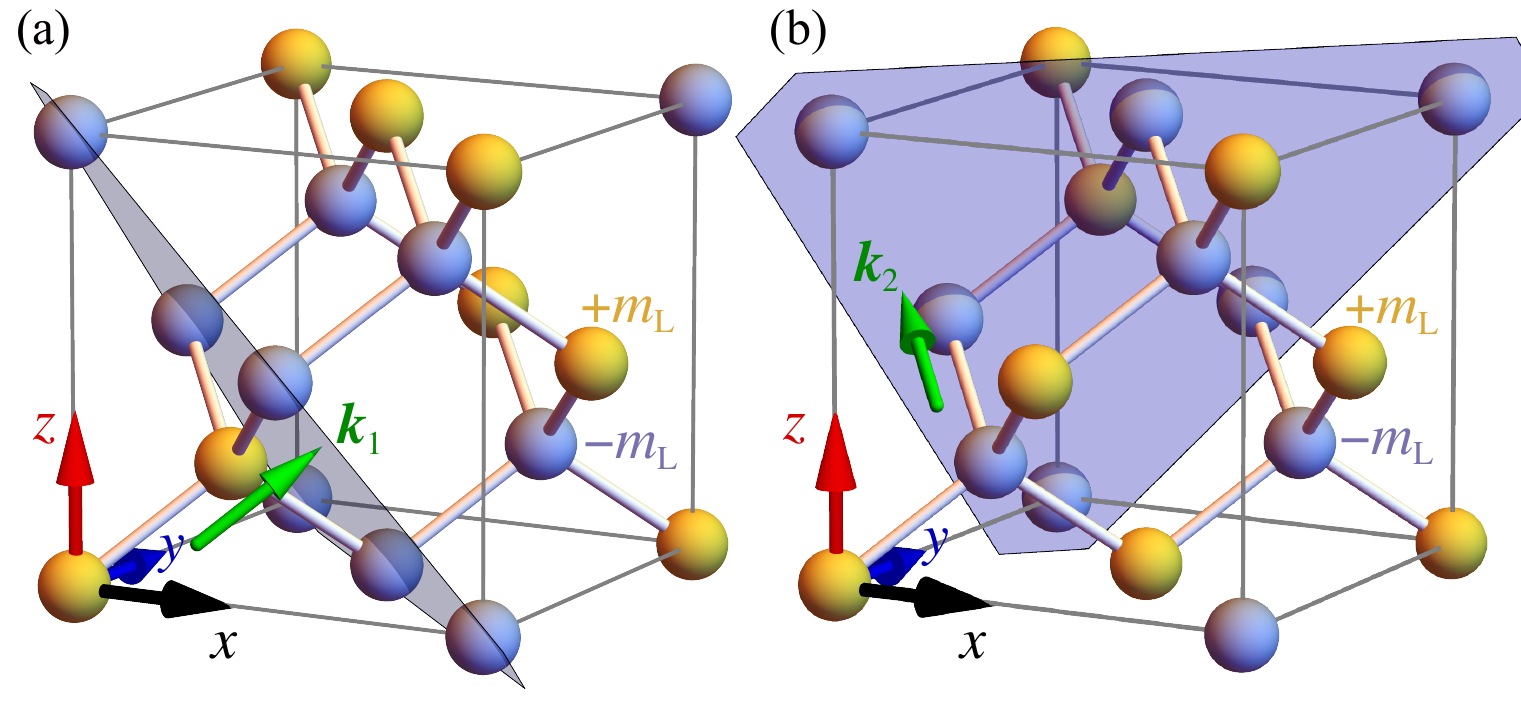}
%\vspace{-10mm}

\end{center}
\caption{Real space sublattice patterns for (a) the single-$\bq$ order with the ordering wavevector (green arrow) $\bq=\bk_1$ and (b) the single-$\bq$ order with $\bq=\bk_2$. Color of spheres represents the order parameter $\pm m_{\rm L}$. A plane on which the order parameters are uniform is emphasized by a shaded cut.}
\label{fig:8}
\end{figure}

First, we show the sublattice order parameter configuration for the single-$\bq$ phases. To avoid representing quadrupolar degrees of freedom by potentially misleading vector arrows, we use colors to distinguish different local configurations in Fig.~\ref{fig:8}, where the real-space sublattice configurations are shown for the single-$\bq$ phases for (a) $\bq=\bk_{1}$ and (b) $\bq=\bk_2$. The color of spheres indicates the sign of  $m_{\rm L}$ in the thermal average of order parameter $\langle \bm{m}_j\rangle =(m_{\Gamma}+m_{\rm L},0)$ or $(m_{\Gamma}-m_{\rm L},0)$ for the 1$\bq$(co) and $\langle\bm{m}_j\rangle=(m_{\Gamma},m_{\rm L})$ or $(m_{\Gamma},-m_{\rm L})$ for the 1$\bq$(ca), where $m_{\Gamma}$ and $m_{{\rm L}}$ denote the corresponding order-parameter amplitudes for the $\Gamma$ and L points, respectively. Note the correspondences: $\langle\bm{m}_j\rangle\propto (1,0)\leftrightarrow O_{20}\sim 2z^2-x^2-y^2$ and  $(0,1)\leftrightarrow  O_{22}\sim \sqrt{3}(x^2-y^2)$. 

%%%%%%%%%%%%%%%%%%%%%%%%%%%%%%%%%%%%%%%%%%%%%%%%%%%%%%%%%%

%%%%%%%%%%%%%%%%%%%%%%%%%%%%%%%%%%%%%%%%%%%%%%%%%%%%%%%%%% 

As represented by the shaded plane perpendicular to $\bk_n$, there are planes where the order parameters are uniform. These planes are arranged along the $\bk_n$ direction forming a layered structure. For sites with $\pm m_{\rm L}$, there are three nearest-neighbor sites with $\mp m_{\rm L}$ and one site with $\pm m_{\rm L}$.  For $\bq=\bk_{1}$, the original unit-cell consists of the two sites with $m_{\rm L}$ (yellow) at the left bottom in Fig.~\ref{fig:8}(a). This corresponds to the factor $p_1(j)$ in Eq.~(\ref{eq:P}). In contrast, for $\bq=\bk_{2}$, the two spheres in the original unit-cell are $m_{\rm L}$ and $-m_{\rm L}$ in Fig.~\ref{fig:8}(b), reflecting the factor $p_2(j)$ in Eq.~(\ref{eq:P}). As clearly visible in Figs.~\ref{fig:8}(a) and \ref{fig:8}(b), the two domains shown are equivalent after a $\pi/2$ rotation along the $z$ axis and an appropriate translation.

\subsection{Double-$q$ orders}
\begin{figure}[t!]
\begin{center}
\includegraphics[width=0.35\textwidth]{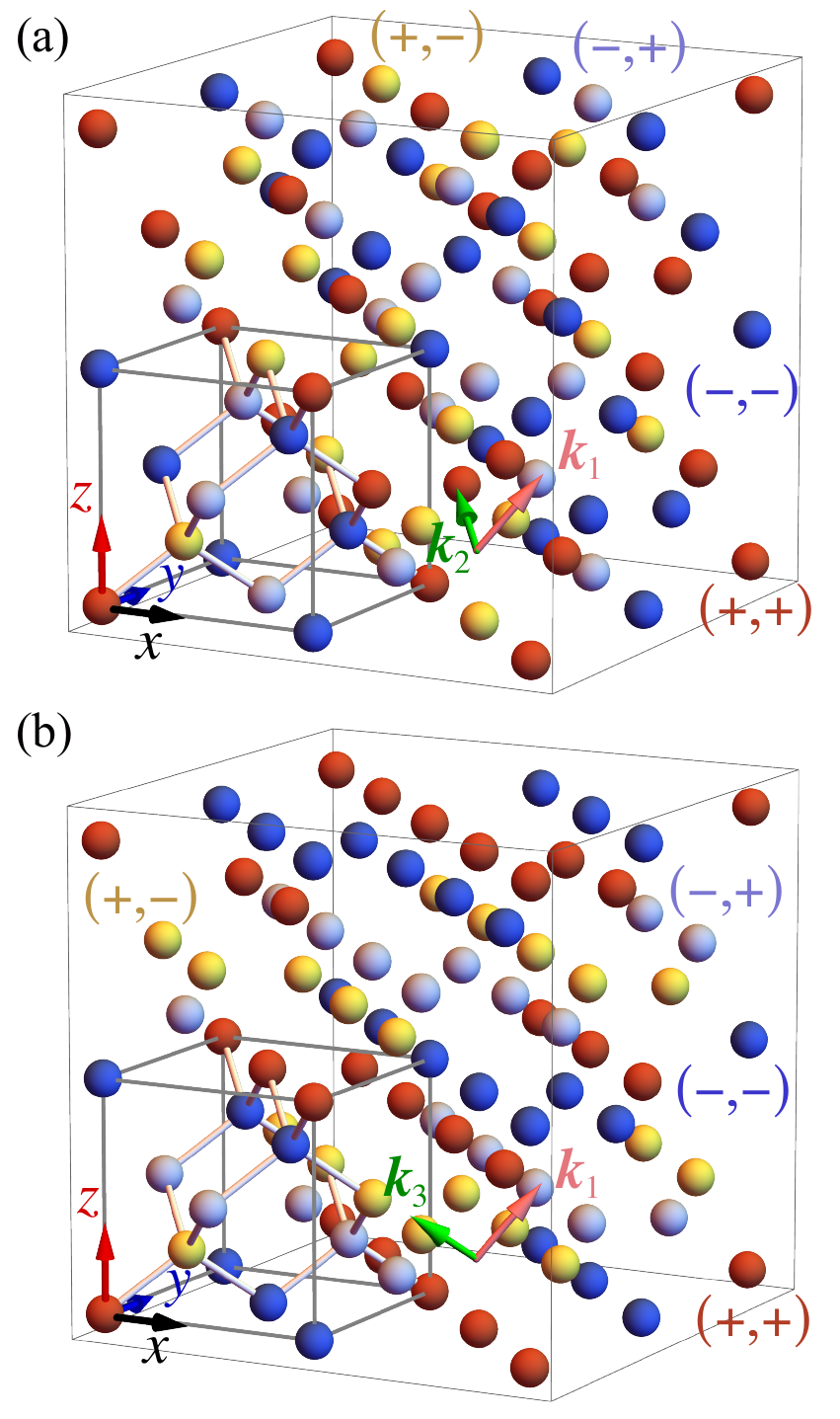}
%\vspace{-2mm}
\end{center}
\caption{Real-space sublattice patterns for (a) the double-$\bq$ order with the ordering wavevectors $\bq=\bk_1$ (orange arrow) and $\bk_2$ (green arrow) and (b) the double-$\bq$ order $\bq=\bk_1$ (orange arrow) and $\bk_3$ (green arrow). 
The four colors distinguish the sign combinations of the two L-point order parameters. The cubic unit cell is shown by a small cubic box. 
}
\label{fig:9}
\end{figure}

Second, the configurations for the double-$\bq$ phases with ${\bm m}_{\rm L}\perp \bm{m}_{\rm L'}$ are shown in Fig.~\ref{fig:9}, where we use the same convention as in Fig.~\ref{fig:8} to illustrate the order parameters. For illustrating the periodicity of the double-$\bq$ structure, we show a region containing eight conventional cubic unit cells and the cubic unit cell is indicated by a thick box in the lower-left corner of each panel.

The four colors distinguish the four possible sign
combinations of the two L-point modes,
$(\sigma_{{\rm L},j},\sigma_{{\rm L}',j})
=(\pm1,\pm1)$.
The thermally averaged local quadrupole moment can be expressed in terms of the uniform and two L-point ordering components as 
$\langle \bm{m}_j\rangle \simeq \bm{m}_\Gamma
+\sigma_{{\rm L},j}\bm{m}_{\rm L}
+\sigma_{{\rm L}',j}\bm{m}_{\rm L'}$. For a representative domain with
$\bm{m}_\Gamma=(m_\Gamma,0)$,
$\bm{m}_{\rm L}=(m_{\rm L},0)$, and
$\bm{m}_{\rm L'}=(0,m_{\rm L'})$,
the orientation of the local quadrupole
moment in quadrupole space is described by
\begin{align}
\tan\eta_j
=\frac{\sigma_{{\rm L}',j}m_{\rm L'}}
{m_\Gamma+\sigma_{{\rm L},j}m_{\rm L}}.
\end{align}
The corresponding local quadrupole is proportional to
$O_{20}\cos\eta_j+O_{22}\sin\eta_j$.

In Fig.~\ref{fig:9}(a) the two ordering wavevectors are $\bk_1$ and $\bk_2$ as indicated by the orange and green  arrows, respectively. Sites of the same color form one-dimensional chains along the $[0\bar{1}1]$direction, which is perpendicular to both $\boldsymbol k_1$ and $\boldsymbol k_2$ and each site has no nearest-neighbor site with the same color, e.g., a yellow sphere is surrounded by two gray spheres, one red sphere, and one blue sphere.  When the ordering wavevectors are switched to $\bk_1$ and $\bk_3$ as shown in Fig.~\ref{fig:9}(b), the direction of one-dimensional chains becomes $[\bar{1}10] \perp \bk_1, \bk_3$. The structure in Fig.~\ref{fig:9}(b) is identical to that in Fig.~\ref{fig:9}(a) after $C_3$ rotation along [111] direction, where $\bk_1\to \bk_1$ and $\bk_3\to -\bk_2\equiv \bk_2$.

%%%%%%%%%%%%%%%%%%%%%%%%%%%%%%%%%%%%%%%%%%%%%%%%%%%%%%%%%%

\bibliographystyle{jpsj}
%\bibliography{references}

\end{document}